# DAEMONS AND DAMA:

# THEIR CELESTIAL-MECHANICS INTERRELATIONS

Edward M. Drobyshevski[1] and Mikhail E. Drobyshevski[1,2]

[1]) *Ioffe Physico-Technical Institute, Russian Academy of Sciences, 194021 St.Petersburg, Russia*
[2]) *Astronomical Department, Faculty of Mathematics and Mechanics, St.Petersburg State University, Peterhof, 198504 St-Petersburg, Russia*

**Abstract.** The assumption of the capture by the Solar System of the electrically charged Planckian DM objects (daemons) from the galactic disk is confirmed by the St.Petersburg (SPb) experiments detecting particles with $V < 30$ km/s. Here the daemon approach is analyzed considering the positive model independent result of the DAMA/NaI experiment in Gran Sasso. The maximum in DAMA signals observed in the May-June period is explained as being associated with the formation behind the Sun of a trail of daemons that the Sun captures into elongated orbits as it moves to the apex. The range of significant 2-6-keV DAMA signals fits well the iodine nuclei elastically knocked out of the NaI(Tl) scintillator by particles falling on the Earth with $V = 30$-$50$ km/s from strongly elongated heliocentric orbits. The half-year periodicity of the slower daemons observed in SPb originates from the transfer of particles, that are deflected through ~90°, into near-Earth orbits each time the particles cross the outer reaches of the Sun which had captured them. Their multi-loop (cross-like) trajectories traverse many times the Earth's orbit in March and September, which increases the probability for the particles to enter near-Earth orbits during this time. Corroboration of celestial mechanics calculations with observations yields ~$10^{-19}$ cm$^2$ for the cross section of daemon interaction with the solar matter.

### I. Introduction. Crisis in the Search for DM Candidates

The origin of dark matter (DM) has intrigued researches for several decades. It has become increasingly clear that these are neither neutrinos with $m \geq 10$ eV nor massive compact halo objects (MACHOs) (Evans and Belokurov, 2005). In the most recent decade, efforts were focused primarily on the search for weakly interacting massive particles (WIMPs) and similar objects with a mass of ~10-100 GeV, whose existence is predicted by some theories of elementary particles beyond the Standard Model. It was believed self-evident from the very beginning that the cross section of their interaction with nucleons, *s*, should be larger than that of their mutual annihilation (~$3\times10^{-34}$ cm$^2$) (Primack *et al*., 1988). The level reached presently is $s \sim 10^{-42}$-$10^{-43}$ cm$^2$, but still no reliable and universally accepted results have been reported. One cannot avoid the impression that the researchers are pursuing an imaginary but nonexisting horizon, in other words, that the WIMPs as they were conceived of originally simply do not exist.

The only data that could possibly be treated as evidence for the existence of WIMPs, or more generally a DM particle component in the galactic halo, were obtained by the DAMA collaboration in their 7-year-long experiment (Bernabei *et al.*, 2003). The evidence is a yearly modulation of the number of 2-6-keV signals accumulated with ~100 kg of NaI(Tl) scintillators. The modulation is ~5% and reaches a maximum some time at the beginning of June; it could be attributed to a seasonal variation of a ground-level flux of objects from the galactic halo caused by the Earth's orbit being inclined with respect to the direction of motion of the Solar system around the galactic centre (Bernabei *et al.*, 2003). The statistical significance of the modulation ($6.3\sigma$) is high enough to leave no doubt in its existence. The experiments being performed on other installations do not, however, support these results. To interpret this situation, the scientists running the DAMA have to consider different types of WIMPs and different modes of their interaction with matter. Recall, for instance, the recent assumption of a light pseudoscalar and scalar DM candidate of ~keV mass (see Bernabei *et al.*, 2006, and refs. herein). Another approach was put forward by Foot (2006). He believes that the DAMA signals originate from the Earth crossing a stream of micrometeorites of mirror matter.

The purpose of the present paper is to show that the effects observed by DAMA/NaI, including the yearly variation of the signal level, allow an interpretation drawn from the St.Petersburg (SPb) experiments on detection of DArk Electric Matter Objects (daemons), which presumably are Planckian elementary black holes carrying a negative electric charge ($m_d \approx 3 \times 10^{-5}$ g, $Ze = -10e$) (Drobyshevski, 1997a,b).

Starting from March 2000, we have been reliably detecting by means of thin spaced ZnS(Ag) scintillators, both in ground-level and underground experiments, signals whose separation corresponds to ~10-15 km/s, i.e., the velocity of objects falling from near Earth, almost circular heliocentric orbits (NEACHOs). The flux is ~$10^{-7}$ cm$^{-2}$s$^{-1}$ and it varies with $P$ = 0.5 yr, to pass through maxima in March and September (Drobyshevski, 2005a; Drobyshevski and Drobyshevski, 2006; Drobyshevski *et al.*, 2003). (Note, that attempts were made to treat these results in terms of possible properties of mirror matter also (Foot and Mitra, 2003), but observations of a daemon flux directed upward, i.e., from under the ground level, are apparently in conflict with this interpretation.)

**II. Specific Features of the Traversal of the Sun by Daemons**

As the Sun moves through the interstellar medium with a velocity $V_\infty$, particles of the latter (we assume for the sake of simplicity that their velocity is $\ll V_\infty$) become focused by gravitation of the Sun, so that its effective cross section becomes $S_{eff} = \pi p_{max}^2 = \pi R_\odot^2 [1+(V_{esc}/V_\infty)^2]$ (Eddington, 1926). Here $R_\odot$ is the radius of the Sun, $V_{esc}$ = 617.7 km/s is the escape velocity from its surface, and $p_{max}$ is the maximum value of the impact parameter (the impact parameter is the distance between the continuation of the $V_\infty$ vector and the center of the Sun). For $V_\infty$ = 20 km/s, $p_{max}$ = 30.9$R_\odot$.

In crossing the Sun with a velocity of ~$10^8$ cm/s, daemons are slowed down. One cannot calculate at present the associated decelerating force, because a negative daemon captures protons and heavy ($Z_n > Z$) nuclei, catalyzes proton fusion reactions, decomposes somehow the nucleons in nuclei etc. As a result, the effective charge of the daemon, complete with the particles captured and carried by it, varies continuously. Straightforward estimates show, however, that daemons of the galactic disk with a velocity dispersion of ~4-30 km/s (Bahcall *et al.*, 1992), are slowed down strongly enough to preclude the escape to infinity of many of them as they pass through the Sun (Drobyshevski, 1996, 1997a).

Such objects move along strongly elongated trajectories with perihelia within the Sun. Subsequent crossings of the Sun's material bring about contraction of the orbits and their escape under the Sun's surface. If, however, a daemon moving on such a trajectory passes through the Earth's gravitational sphere of action, it is deflected, which will result in the perihelion of its orbit leaving the Sun with a high probability. The daemon will be injected into a stable, strongly elongated Earth crossing heliocentric orbit (SEECHO). Straightforward estimates made in the gas kinetic approximation and using the concepts of mean free path length etc. suggest that daemons build up on SEECHOs to produce an Earth crossing flux of ~$3 \times 10^{-7}$ cm$^{-2}$s$^{-1}$ (Drobyshevski, 1997a). (These were fairly optimistic calculations performed for a rough estimation of the parameters of the daemon detector that was being designed at that time, in 1996.)

In subsequent inevitable crossings by SEECHO daemons of the Earth's sphere of action, their orbits deform to approach that of the Earth; these are near-Earth almost circular heliocentric orbits (NEACHOs). And whereas daemons moving in nearly parabolic SEECHOs strike the Earth with velocities of up to $V_\oplus \sqrt{3} = 51.6$ km/s (here $V_\oplus = 29.79$ km/s is the orbital velocity of the Earth), the NEACHO objects fall on the Earth with a velocity of only ~10(11.2)-15 km/s.

Estimates of the ground-level SEECHO daemon flux made in 1996 were based on simple concepts of an isotropic flux of galactic disk daemons incident on the Sun. Our subsequent experiments that demonstrated a half-year variation made it clear that the flux is not isotropic, probably because of the motion of the Solar system relative to the DM population of the disk. We know now also that the daemons detected by us fall, judging from their velocity, from NEACHOs (Drobyshevski, 2005b; Drobyshevski *et al.*, 2003).

**III. Calculation of the Passage of Daemons through the Sun**

It appears only natural to assume the flux variations with $P = 0.5$ yr to be a consequence of the composition of the Earth's orbital motion around the Sun and of the Sun itself relative to the galactic disk population.

The Sun moves relative to the nearest star population with a velocity of 19.7 km/s in the direction of the apex with the coordinates $A = 271°$ and $D = +30°$ (equatorial coordinates) or $L" = 57°$ and $B" = +22°$ (galactic coordinates) (Allen, 1973).

Initially, rather than delving into the fundamental essence of the processes underlying the celestial mechanics, we invoked a simplified concept of a "shadow", which is produced by daemons captured into SEECHOs from the galactic disk by the moving Sun, and of the corresponding "antishadow" created by some daemons crossing the Sun in the opposite direction (i.e., in the direction of its motion), an approach that had been reflected in our earlier publications (Drobyshevski, 2004; Drobyshevski *et al.*, 2003).

In a new approach to calculation of the passage of objects through the Sun we made use of the celestial mechanics integrator of Everhart (1974). It was adapted to FORTRAN by S.Tarasevich (Institute of Theoretical Astronomy of RAS) for use with the BESM-6 computer. In ITA (and now in the Institute of Applied Astronomy of RAS) it was employed in calculation of asteroid ephemeredes and precise prediction of the apparition of comets allowing for the action of known planets. We made two important refinements on the code, more specifically, we (*i*) introduced the resistance of the medium in the simplest gas dynamics form, $F = \sigma \rho V^2$ (Drobyshevski, 1996) (where $\sigma$ is the effective cross section of a particle, and $\rho$ is the medium density) and (*ii*) took into account that the Sun is not a point object but has instead a density distributed over the volume (we used the model of the Sun from Allen (1973)).

The very first calculations revealed that the trajectories of particles falling on the Sun and crossing it have a non-closed, many-loop pattern (Drobyshevski, 2005b). This should certainly have been expected, because inside the Sun particles move in a gravitational field not of a point but rather of a radius-dependent mass, so that the trajectories do not close and form instead a rosette, whose petals appear successively in the direction opposite to that of a body moving around the Sun (see, e.g., Figs.4 and 5 below).

This prompted us to consider the possibility of combining and explaining the results of DAMA/NaI and of our experiments in terms of a common daemon paradigm, all the more so because earlier attempts (Drobyshevski, 2005a) had succeeded in proposing an interpretation of the so-called "Troitsk anomaly", i.e., a displacement of the tritium β-spectrum tail occurring with a half-year periodicity.

This approach:
a) would hopefully provide an answer as to why the results of the DAMA/NaI are not confirmed by other WIMP experiments;
b) would permit us to understand why the intensity of the scintillation signals assigned to recoil nuclei lies in the 2-6-keV interval (here 2 keV is the sensitivity threshold of the DAMANaI detector) and not higher, whereas elastic interaction of nuclei with WIMPs of the galactic halo ($V$ = 200-300 km/s) should seemingly produce signals with energies of up to ~200 keV.

**IV. On How to Corroborate the St.Petersburg and DAMA Experiments**

The first question that comes immediately to mind is how could one explain the twofold difference in the signal periodicity between the SPb and DAMA experiments?

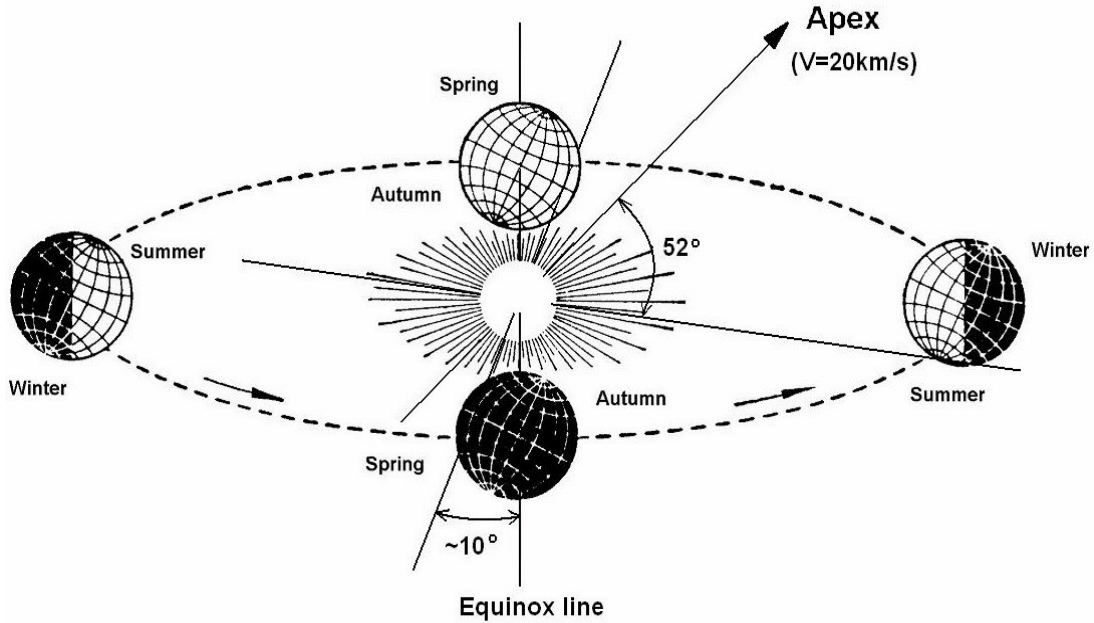

Figure 1. Scheme of motion of the Sun and the Earth towards the apex (see text).

Let us begin with the SPb experiment.
Figure 1 shows schematically the motion of the Sun together with the Earth in the direction of the apex. The angle between the plane of the Earth's orbit, the ecliptic, and the apex direction is approximately $\alpha_0 = 52°$-$53°$ (the direction to the apex, just as the velocity $V_\infty$, depends on the stars (or interstellar gas clouds etc.) one chooses as references (Allen, 1973); we assume in what follows $\alpha_0 = 52°$ and $V_\infty = 20$ km/s). We assume also the angle between the straight line lying in the ecliptic plane and normal to the apex direction and the equinox line to be about 10°; as the Earth moves along its orbit, it crosses first this line, and after that, the equinox line. This order of crossing fits our measurements of the positions of the maxima in the primary daemon flux (Drobyshevski, 2005a; Drobyshevski and Drobyshevski, 2006), which occur some time in the first decade of March or September (and incidentally coincides with $A = 258°$ for the Solar apex relative the interstellar gas).

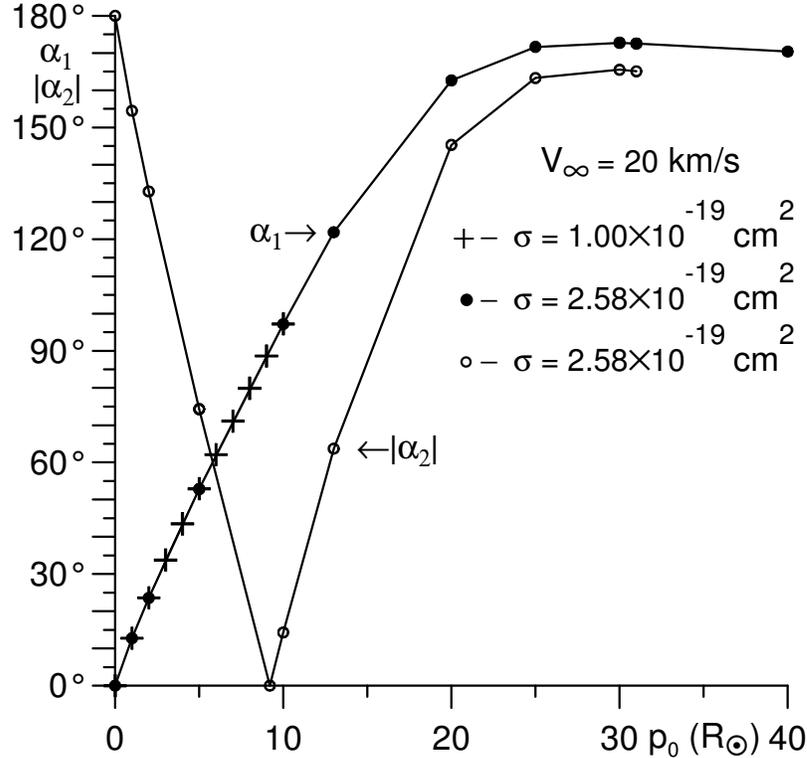

Figure 2. Angle of deviation for an object having passed through the Sun depending on the impact parameter $p_0$ at different cross-sections $\sigma$ of its interaction with matter (at the object mass $3\times10^{-5}$ g). $\alpha_1$ is the angle of deviation from the initial velocity $V_\infty$ direction after the first passage through the Sun; $\alpha_2$ is the angle of deviation from the same direction after the second passage.

Figure 2 plots the angle of deviation $\alpha_1$ of a material particle from the direction of its initial velocity $V_\infty$ after emergence from the Sun again to infinity or to the aphelion of the first loop (i.e. at $R_1$) of its trajectory vs. the impact parameter $p_0$ (the dependence of $R_1$ on the impact parameter $p_0$ is given in Fig.3). We consider subsequently only cases with $p_0 < p_{max}$. Interestingly, in the case of multi-loop trajectories, which is possible for $\sigma \ne 0$, angular deflections of subsequent from preceding loops differ little from $\alpha_1$, although they gradually decrease. The value of $\sigma$ can be estimated from a comparison of further calculations with experiment.

Straightforward reasoning suggests that the two maxima observed in March and September should be a consequence of passage through the Earth's orbit of daemons with an impact parameter about $p_0 = \pm 9.162 R_\odot$, where they are deflected by the Sun through $\alpha_1 \approx 90°$ to either side. Moreover, the presence of these maxima suggests also that they originate from the daemons that had already been captured by the Sun for $\sigma \geq 0.78 \times 10^{-19}$ cm$^2$ but return repeatedly to it and cross its body. Figure 4 (with a table) provides an idea of such trajectories.

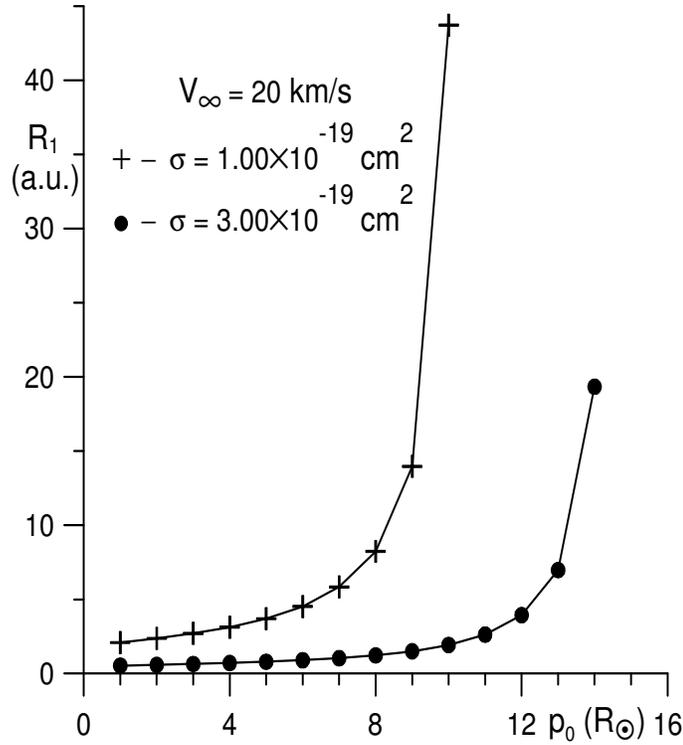

Figure 3. Maximum distance $R_1$ the object reaches after the first passage through the Sun versus $\sigma$ and $p_0$.

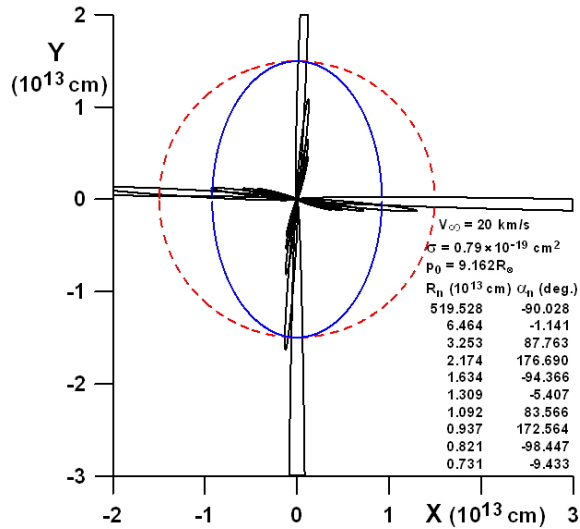

Figure 4. An example of multi-loop (cross-like) trajectory of an object being braked by the Solar matter (the Sun's center is at $X = 0$, $Y = 0$) for repeated passages through the Sun's body. Object of $3\times10^{-5}$ g mass and cross-section $\sigma = 0.79\times10^{-19}$ cm$^2$ falls from infinity ($X = -\infty$; $V_\infty = 20$ km/s) with an impact parameter $p_0 = Y(-\infty) = 9.162 R_\odot$. The figure plane contains the direction to the apex and the normal to it lying in the plane of the ecliptic. The figure shows also an ellipse with the major semi-axis of 1 AU, i.e., the projection of the Earth's orbit (the dotted circle of 1 AU radius is given as a scale for the reader's orientation).

A particle moves along a trajectory making a right cross. First, in traversing the Sun, it is deflected through 90° and, in crossing the Earth's orbit, escapes with $\sigma = 0.78 \times 10^{-19}$ cm$^2$ to $R_1 \rightarrow \infty$ (calculations for Fig.4 are made for $\sigma = 0.79 \times 10^{-19}$ cm$^2$ to avoid calculations with too great a value of $R_1$). Thereafter, returning, now from outside, it crosses for the second time the Earth's orbit and, hitting the Sun on completion of the first loop, it again deflected, leaving it through nearly the right angle. Now the daemon moves along the petal oriented in the anti-apex direction. But here, although $R_2 > 1$ AU, it does not cross the Earth's orbit because of the large inclination of the ecliptic. The subsequent two crossings of the Earth's orbit (here still $R_3 > 1$ AU) from the side opposite to that of the first transit, are completed by the daemon after the return and the third crossing of the Sun. In making the fourth passage through the Sun after the return, the particle moves toward the apex. Here, depending on the value of $\sigma$ and completing the cross, the daemon can move away from the Sun to a distance $R_4 > 1$ AU (but here again, because of the inclination of the ecliptic to the apex direction, it does not cross the Earth's orbit), but may not reach $R_4 = 1$ AU at all. The first value of $R_4 > 1$ AU corresponds to $\sigma_1 = 0.78 \times 10^{-19}$ cm$^2$ at which the resistance of the solar material in the first passage was just high enough to absorb the excess energy $\Delta E = m_d V_\infty^2/2$, i.e., the particle was captured by the Sun. The second (upper) value $\sigma_3 = 1.415 \times 10^{-19}$ cm$^2$ (for the minimum value $R_3 = 1$ AU) can be estimated under the assumption that on its third passage the daemon can finally reach and crosses slightly the Earth's orbit (i.e., $R_3 > 1$ AU). The validity of the latter assumption is argued for at least by our observation in autumn and spring of two distinct maxima; the fourfold (or even six-fold - see below) crossing by daemons of the Earth's orbit increases, accordingly, the probability of their transfer from the loop trajectories into SEECHOs and, subsequently, in NEACHOs, whence they fall on the Earth with $V \approx$ 10-15 km/s. Thus, we arrive at $0.78 \times 10^{-19} \leq \sigma < 1.415 \times 10^{-19}$ cm$^2$ (we point out once more that the values of $\sigma$ thus found depend on the accepted daemon mass; they should be proportional to $m_d$). For $\sigma = \sigma_3$, the Sun does not capture the daemon at $p_0 > 12.01 R_\odot$, i.e., when the daemon initially passes through more rarefied outer layers of the Sun it moves to infinity again.

To choose a still more optimistic scenario, assume a daemon that for $\sigma_1 = 0.78(0.79) \times 10^{-19}$ cm$^2$ crosses the Earth's orbit in the fifth loop as well (with $R_5 = 1.0919 > 1$ AU). The condition $R_5 = 1$ AU yields $\sigma_5 = 0.849 \times 10^{-19}$ cm$^2$. With $\sigma_1 < \sigma < \sigma_5$, the daemon has now crossed the Earth's orbit six times! In this respect, Jupiter is markedly behind the Earth (with only two crossings), while Venus is on the winning side (8 crossings).

Figure 3 presenting the $R_1(p_0)$ relation in a graphic form for $\sigma = 1 \times 10^{-19}$ and $3 \times 10^{-19}$ cm$^2$ facilitates estimation of the energy losses suffered by the daemon in traversing the Sun, of the number of such traversals etc.

Turning now back to the DAMA/NaI experiment, the daemon can obviously cross the set-up in June provided it falls after the first traversal of the Sun in the plane of the ecliptic, i.e., if it is deflected through $\alpha_1 = 52°$, which occurs at $p_0 = 4.90 R_\odot$. The above estimates of $\sigma$ suggest (see Fig. 5a) that $R_1 > 1$ AU, and that the second loop extends up to $R_2 > 1$ AU also, while naturally not crossing the Earth's orbit, because it leaves the ecliptic plane. In June, the second loops of the trajectories with $p_0 = -6.26 R_\odot$ enter the plane of the ecliptic as well and extend in it up to $R_2 \approx 2 > 1$ AU (Figs.5c,d). In December, the second loops of trajectories with $p_0 = 2.215 R_\odot$ enter the ecliptic plane (Fig.5b). At $\sigma = \sigma_5 = 0.849 \times 10^{-19}$ cm$^2$ they have $R_2 = 1.14$ AU, i.e. they are able to cause SEECHOs in December, and only at $\sigma \geq 0.94 \times 10^{-19}$ cm$^2$ $R_2$ becomes $\leq 1$ AU.

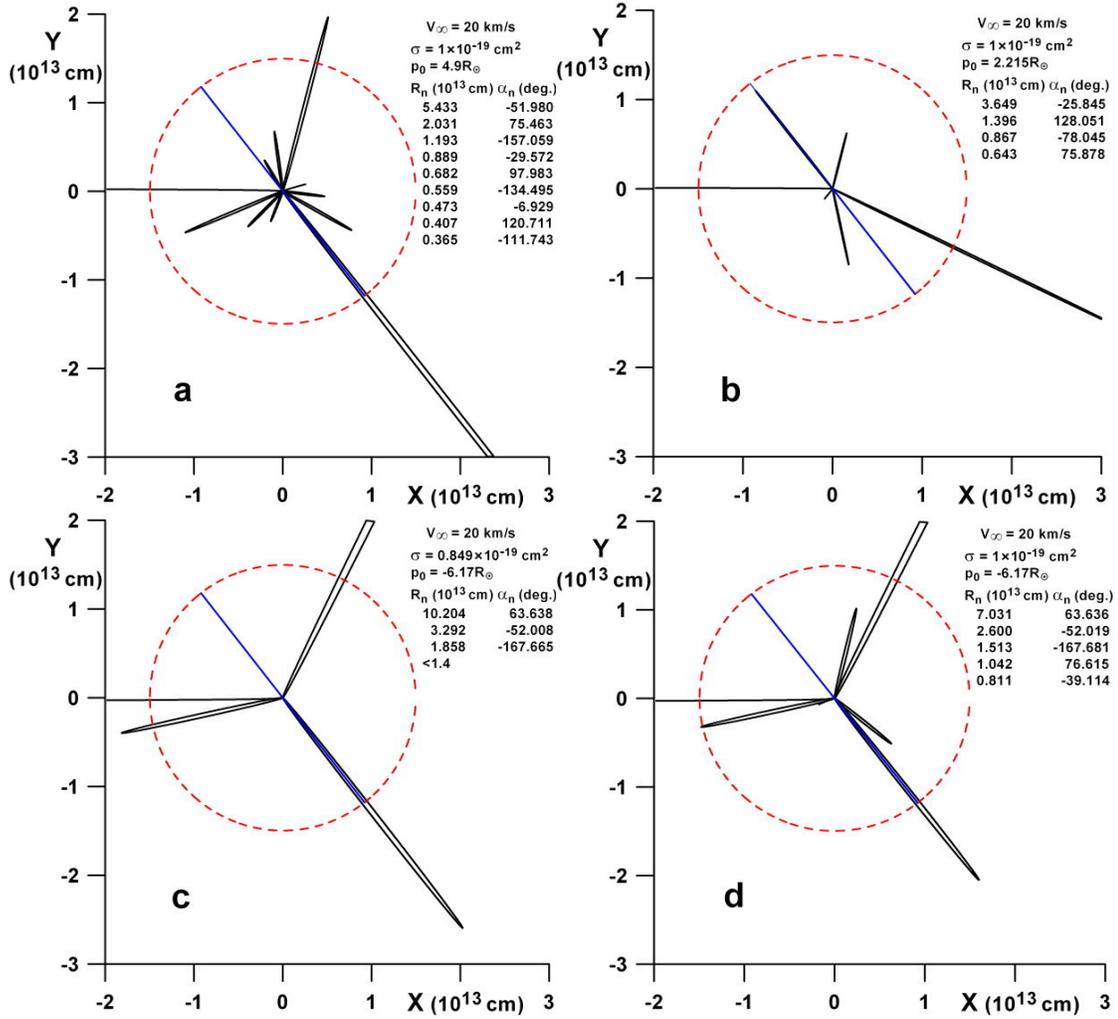

Fig. 5. The same as Fig.4, but here the Earth's orbit is projected into the figure plane as a straight line segment of 2 AU length with 52° inclination to the apex direction (December is up-left, June is down-right).

Thus the calculations performed for $\sigma = \sigma_1 = 0.78\times10^{-19}$ cm$^2$, $\sigma = \sigma_5 = 0.849\times10^{-19}$ cm$^2$, $\sigma = 1\times10^{-19}$ cm$^2$, and $\sigma = \sigma_3 = 1.415\times10^{-19}$ cm$^2$ suggest that the daemons captured in traversing the Sun produce behind it a fairly smeared trail ("shadow") through which the Earth passes in May-June-July, but which, generally speaking, does not reach the part of the Earth's orbit oriented in the direction of the apex and corresponding approximately to the November-January period. This is easy to understand, because the second loops of the trajectories which fall into the apex hemisphere and could produce an "antishadow" correspond to small $p_0$, i.e., to particles passing through the dense central part of the Sun, where they suffer the strongest deceleration. This is why, in particular, the second loop of the trajectory of the daemon that crossed the Sun at $p_0 = 2.215 R_\odot$ and fell into the ecliptic plane exactly in December, simply cannot reach the Earth for $\sigma \geq 1\times10^{-19}$ cm$^2$ (Fig.5b). It is thus clear that the ground level flux of daemons from SEECHOs should exhibit a distinct 1-year periodicity with a minimum some time in December.

Superimposed on this is the half-year wave of the NEACHO objects, which appear as a result of having transferred from numerous SEECHOs just in periods before the equinoxes (note these SEECHOs are realizing at both signs of $p_0 = \pm 9.162 R_\odot$). Such transitions are more probable in March and September because of the appearance of a noticeably larger number of objects in SEECHOs with comparatively short semimajor axes lying close to these ecliptic plane zones. The daemons entering these SEECHOs come from the cross-shaped rosette trajectories, along which the same object crosses the Earth's orbit twice (or even thrice) back and forth, the second (and all the more so, the third) time doing it <u>with a noticeably below the parabolic velocity</u>. Significantly, the projection of the SEECHO object velocity vector on the Earth's orbit reaches its maximum values here, with the correspondingly increasing duration (and efficiency) of the Earth's gravitational perturbations. Note also that the ratio of the minor to major semi-axes for the SEECHOs produced in the capture of objects that had crossed the outer zones of the Sun ($p_0 \approx 10 R_\odot$) exceeds those for the objects with $p_0 \to 0$ (compare Figs.4 and 5), which, on the whole, also acts so as to increase the velocity vector projection on the Earth's orbit.

**V. The Manifestation of Daemons in the SPb and DAMA/NaI Detectors**

Our SPb detector, made up of thin spaced ZnS(Ag) scintillators, was sensitive to the passage of only fairly low-velocity (<30 km/s) daemons. The reason for this, we believe presently, is that successive "disintegrations" of daemon-containing nucleons in the Zn (or Fe) nucleus captured by the daemon occur with an interval of ~$10^{-6}$ s, whereas the characteristic dimension of the set-up is ~20-30 cm. At higher velocities, the complex consisting of the captured nucleus (and, possibly, a cluster of atoms) and the daemon traverses the system carrying an excessive positive charge, which is readily compensated by electrons captured on the way, and, therefore, the daemon does not interact with new nuclei with an attendant generation of a noticeable secondary signal.

The DAMA/NaI experiment with a ~100 kg NaI(Tl) scintillators was designed for measurement of the annual modulation signature. In case of WIMPs the measured quantity is the energy of the recoil nuclei knocked out by heavy (~10-100 GeV) WIMPs of the galactic halo. The set-up is thought to be sensitive to interactions occurring both with I and Na.

Note that the sensitivity threshold of the system, ~2 keV, corresponds to the velocity of an iodine nucleus of 55 km/s if one takes the quenching factor to be about unity (the quenching factor is a ratio of efficiencies of producing scintillations by particles under consideration and electrons of the same energy). Assuming elastic interaction with a very massive particle, the latter, to produce a 2-keV signal, should move with a velocity of ~30 km/s (in an elastic head-on collision, the velocity of the light particle, a nucleus, should be twice that of the heavy projectile particle). If the signals are due to WIMPs of the galactic halo ($V_\infty$ = 200-300 km/s), the recoil energy of the iodine nucleus could, seemingly, reach as high as 110-240 keV.

Information on the yearly variation of the flux of particles traversing the DAMA/NaI is provided primarily by signals in the 2-6-keV range (Bernabei *et al.*, 2003, 2006). The 6-keV signal corresponds to the velocity of an elastically colliding projectile of ~ 47 km/s. This figure is in good agreement with the velocity of 51.6 km/s with which a particle in a quasi-parabolic orbit hits the Earth (29.78√3 = 51.6 km/s; this velocity would produce a recoil nucleus with an energy of 7 keV, but allowing for the statistics of other than head-on collisions we would obtain 5-6 keV). But it is with these velocities (when particle energies differ exactly by a factor of three; compare, on the other hand, the 6 and 2 keV which one measures!) that SEECHO daemons fall on the Earth. A truly remarkable coincidence indeed.

A number of additional questions, however, immediately arise here, to which one cannot yet supply unambiguous answers.

Indeed, estimates of the velocity with which daemons escape from geocentric Earth-crossing orbits (GESCO) into the Earth suggest that the resistance offered by a metal-like solid to a daemon moving with a velocity of ~10 km/s is ~$10^{-5}$ dyne (Drobyshevski, 2004), which entails a release of thermal energy of ~6 MeV/cm. It is unclear what energy would be liberated in a dielectric (without conductive electrons) scintillator. If it is heat, the scintillator will not detect it. (On the other hand, it is too high for the cryogenic systems of the type CDMS-I and Edelweiss-I designed for the detection of WIMPs either; see refs. in Bernabei *et al.* (2003).)

The situation is not yet clear with regard to the quenching factor, which we considered above to be about unity for the low-energy iodine nuclei. Neutron elastic scattering experiments give a value of about 0.09 for I and 0.30 for Na (see Bernabei *et al.*, 2003, and refs. therein). We will not give details of such calibrations, nor discuss the different possibilities here (see, however, arXiv:0706.3095).

Also, one should not forget that the DAMA/NaI system, by the multiple hit rejection criterion by Bernabei *et al.* (2003, see Sec.3.3), rejects events with signals appearing simultaneously in two or more scintillators (nine scintillators altogether). But then how (and with what efficiency) do recoil nuclei form in one detector piece only?

One may recall that immediately after leaving a solid, the daemon moves in vacuum (or in air) together with a cluster of atoms, in one of whose nuclei it resides (Drobyshevski, 2005a). When it enters a solid object again, the daemon leaves a larger part of this cluster close to the surface of the object, and moves inside it only with a small part of the cluster, or even only with the remainder of the nucleus in which it rests. It is unclear how efficiently such a complex can initiate scintillations. It is conceivable that as long as it carries an excess positive charge, it is surrounded by electrons, and, in moving as a conventional heavy atom (or ion with $Z_i = 1$) with a relatively low velocity (<50 km/s; recall that this corresponds to less than ~2 keV energy of an iodine nucleus), but in a rectilinear trajectory and without noticeable deceleration, through the dielectric and only moving atoms apart rather than penetrating into them, it will excite only phonons but not scintillations. The daemon resides in this state for tens of microseconds, "digesting" gradually the nucleons of the nucleus it is carrying and traveling tens of cm in this time. (We are not discussing here the points bearing on possible modes of "digestion" by the daemon, an elementary black hole, of nucleons that could leave no trace in a scintillation detector.) Eventually, however, the daemon/nuclear-remainder complex acquires zero charge to become a particle of the neutron type. This state lasts until the next proton in the nucleus disintegrates and the system then acquires a negative charge, ~$10^{-6}$ s (with the distance traveled ~5 cm). It is in this state that this neutral (but supermassive, ~$3\times10^{-5}$ g) complex, 1-3 fermi in size, passes <u>through</u> electronic shells of atoms and is capable, in a path length of ~5 cm, to produce a recoil nucleus with a double velocity of up to ~100 km/s. It is such SEECHO-daemon-caused events that satisfy the multiple hit rejection criterion (Bernabei *et al.*, 2003) that are possibly detected by the DAMA/NaI. The processes involved here certainly need a deeper analysis.

## VI. Conclusions

The daemon approach had offered an explanation for the ~5-eV drift of the tritium β-spectrum tail with a half-year period, the so-called "Troitsk anomaly", and some predictions regarding the KATRIN experiment on direct measurement of the neutrino mass (Drobyshevski, 2005a).

It now appears that one can corroborate within the daemon paradigm the results of the DAMA/NaI experiment with the inferences drawn from the SPb study, which, in addition to detecting daemon populations of different velocities captured by the Solar system and moving within it in orbits of different, including geocentric, populations, established also a half-year variation in the NEACHO flux, demonstrating the advantages of vacuum systems in daemon detection, and revealing some of their remarkable properties and specific features of their interaction with matter.

We find it particularly impressive that the range (2-6 keV) of the recorded signals in which the DAMA/NaI exhibits a yearly periodicity coincides with exactly the same level (2-7 keV) that follows from the celestial mechanics scenario. On the other hand, a more careful analysis of the evolution of daemons captured by the Sun from the galactic disk and governed by celestial mechanics sheds light on reasons underlying the detection of the yearly periodicity of the high-velocity population (30-50 km/s) measured by DAMA, and of the half-year periodicity of the low-velocity population (5-10-30 km/s) in the SPb experiment (in the latter case, the part played by the cross-like multi-loop trajectories of daemons traversing the Sun and captured by it appears significant). If performed with good statistics, the more advanced measurements on DAMA/LIBRA will hopefully also reveal the half-year harmonic in low signal level events (~2-3 keV). Such events originate from the fall of daemons from "short" SEECHOs in March and September.

An analysis of the conditions favouring capture of daemons by the Sun and corroboration of their subsequent possible celestial mechanics evolution with the results gained in the SPb and DAMA/NaI experiments permits one to impose fairly strong constraints (one would almost say, to measure) on the effective cross section $\sigma$ of daemon interaction with the Solar material. It was found to be $0.78 \times 10^{-19} \leq \sigma < 1.4 \times 10^{-19}$ cm$^2$. This is ~500 times the cross section of the neutral "antineon" atom formed by a daemon ($Ze = -10e$) and ten protons it captured, while being 3000-5000 times smaller than the cross section for a daemon/heavy-nucleus complex with electrons surrounding it. On the other hand, $\sigma$ can be governed also by the Coulomb interaction of the daemon changing continuously its effective charge with particles of the solar plasma (Drobyshevski, 1996), which, in turn, may be helpful in refining our knowledge of the characteristics of the solar material.

Obviously enough, the problems addressed in the paper (interaction of daemons both with the solar matter and with the scintillator material, the celestial mechanics and statistical evolution of their ensemble after capture by the Solar system, the part played by the initial conditions and the starting velocity dispersion in the daemon population of the galactic disk, refinement of the apex relative to this population, - it seems it is closer to the apex relative the interstellar gas, not stars (see. Fig.1), transfer to SEECHOs and, subsequently, to NEACHOs and GESCOs etc.) would require a much more careful and comprehensive analysis. This would permit a quantitative comparison of theoretical predictions with future experimental results.